\pdfoutput=1
\documentclass[11pt]{article}

\usepackage[preprint]{acl}

\usepackage{times}
\usepackage{latexsym}
\usepackage{listings}
\usepackage[T1]{fontenc}
\usepackage[utf8]{inputenc}

\usepackage{microtype}

\usepackage{inconsolata}

\usepackage{graphicx}

\usepackage{algorithm} 
\usepackage{algpseudocode} 
\usepackage{adjustbox}
\usepackage{geometry}
\usepackage{comment}
\usepackage{array}
\usepackage{graphicx} 
\algdef{SE}[SUBALG]{Indent}{EndIndent}{}{\algorithmicend\ }%
\algtext*{Indent}
\algtext*{EndIndent}
\usepackage{soul}

\title{AIGCodeSet: A New Annotated Dataset for AI Generated Code Detection}

\author{Basak Demirok\\
  TOBB University of Economics and Technology
  \\
  g.demirok@etu.edu.tr \\\And
  Mucahid Kutlu \\
  Qatar University \\
  mucahidkutlu@qu.edu.qa \\
  }

\begin{document}
\maketitle

\begin{abstract}

While large language models provide significant convenience for software development, they can  lead to ethical issues in job interviews and student assignments. Therefore, determining whether a piece of code is written by a human or generated by an artificial intelligence (AI) model is a critical issue. In this study, we present AIGCodeSet, which consists of 2.828 AI-generated and 4.755 human-written Python codes, created using CodeLlama 34B, Codestral 22B, and Gemini 1.5 Flash. In addition, we share the results of our experiments conducted with baseline detection methods. Our experiments show that a Bayesian classifier outperforms the other models.
\end{abstract}

% En son yazılacak
\section{Introduction}
Artificial Intelligence (AI)-powered large language models are playing an important role in software development, significantly enhancing productivity. Beyond general-purpose chat-based models, specialized models such as CodeLlama, Codestral, and StarCoder have been specifically trained for code generation, enabling more efficient and automated software engineering workflows.

However, the widespread adoption of AI-generated code also introduces critical challenges, particularly concerning reliability and trustworthiness \cite{wang2024your, rabbi2024ai, ambati2024navigating}. These codes can sometimes contain unoptimized algorithms, inefficient structures, or faulty logic.

On the other hand, ethical violations such as candidates trying to deceive employers by using AI tools in job interviews\footnote{https://fortune.com/2024/10/23/elon-musks-xai-cofounder-calls-out-cheating-interviewee-and-now-employers-are-outing-the-ai-tools-being-abused-by-savvy-job-seekers/} and students having AI models do their homework are also a major concern\cite{deng2024promoting}.

% wang2024your = really_safe
% ai_writes = rabbi2024ai
%navigating = ambati2024navigating
%e1 = deng2024promoting

For all these reasons, distinguishing AI-generated codes from human-written codes has become a critical requirement for software security and ethical standards.

In this study, we create a dataset to analyze AI-generated codes and compare them with human-written codes. Within the scope of our study, we compiled human-written codes for selected problems from the IBM CodeNet dataset and generated Python codes using three different models (CodeLlama, Codestral, and Gemini) in three different AI model usage scenarios in response to the same problems.

We perform various preprocessing steps on the obtained dataset to ensure the correctness and consistency of the codes. Then, we analyze structural features such as code length, comment lines, and function definitions to show whether the AI-generated codes exhibit certain patterns compared to human-written codes. Finally, we measure the performance of various basic methods such as Ada vectors, TF-IDF vectors, and Bayesian classifier to distinguish between human-written and model-generated codes.

Our experimental results show that the Bayesian classifier provides the highest sensitivity overall. However, we observe that the detection success of the models varies significantly depending on the major language model and usage scenario used.

The main contributions of our work can be summarized as follows:
\begin{enumerate}

    \item We have created a comprehensive dataset containing a total of 4,755 human-written and 2,828 AI-generated codes that can be used in the detection of AI-generated codes. We share our dataset and codes to support research in this field and to reproduce our experimental results. \footnote{https://huggingface.co/datasets/basakdemirok/AIGCodeSet}

    \item We measured the baseline performances of various machine learning models on our dataset.

\end{enumerate}

\section{Related Work}\label{sec_rel}

Studies on AI-generated codes vary in terms of the target programming language, the purpose of creating the dataset, the size of the dataset, and the models selected and the source from which the human data is collected. For example, in terms of programming language, there are studies on various languages such as Python \cite{idialu2024whodunit, oedingen2024chatgpt, pan2024assessing}, Java \cite{suh2024empirical, pham2024magecode}, Ruby \cite{xu2025codevision}, Go \cite{gurioli2024you}, and PHP \cite{wang2023evaluating}.

% idaliu = idialu2024whodunit
% oedingen2024chatgpt = 
%hung = pan2024assessing
% suh = suh2024empirical
%pham = pham2024magecode
%vision = xu2025codevision
%guri = gurioli2024you
%wang = wang2023evaluating
% xu = xu2024detecting

%Veri kümesi oluşturulma amacı bazı çalışmalarda YZ tarafından üretilen kodların tespit modelini geliştirmek \cite{suh, pham, vision, guri} iken bazı çalışmalar da mevcut YZ içeriğini tespit eden yaklaşımların kod kapsamında çalışıp çalışmadığına bakıyor \cite{wang, hung}.

Datasets for detecting AI-generated codes also require human-written codes. Similar to our work, studies have generally found problem sets defined from different sources and a set of human-written codes that are solutions to these problems, and then had various LLMs generate codes for this problem.
Studies in the literature have obtained human-written codes from various sources such as CodeChef \cite{idialu2024whodunit}, Kaggle, Quescol \cite{pan2024assessing}, CodeNet \cite{xu2025codevision, xu2024detecting} and Rosetta Code project \cite{gurioli2024you}. In our study, we also obtained human-written codes from CodeNet.

Moreover, GPT-based models such as ChatGPT \cite{pan2024assessing, suh2024empirical} and GPT-4 \cite{suh2024empirical, pham2024magecode, xu2025codevision} are widely utilized in the literature. Additionally, models like StarCoder \cite{idialu2024whodunit, gurioli2024you}, CodeBision \cite{pham2024magecode}, and Gemini \cite{suh2024empirical} have also been employed in various studies. In our research, we utilize the CodeLlama, Codestral, and Gemini models.

Studies in the literature \cite{pan2024assessing, oedingen2024chatgpt} have focused on detecting code generated from a given problem. However, in this study, in addition to this problem, we also considered the cases of detecting runtime errors or codes that produce incorrect outputs, which are fixed using AI.

\begin{table*}[!htb]

\centering
  \begin{tabular}{ | p{3cm} |  p{12cm} |}
    \hline
    \textbf{Scenario} & \textbf{Prompt} \\ \hline
    Generating code from scratch &  You are a helpful code assistant. Your language of choice is Python. Don’t explain the code, do not add ``python or '', just generate the code block itself. Here is the problem description: $<$description$>$  \\ \hline
    fixing code that results in a runtime error &  You are an expert Python programmer that  helps  to fix the code for runtime errors. I will give you  first the problem description, then the code that has the error.  You will fix the Python code, just return the Python code block itself without explanations. Remember, do not give any explanations. Here is the problem description:  $<$description$>$ Here is the code  that has the error: $<$code$>$ \\ \hline 
    correcting code with incorrect output  &  You are an expert Python programmer that helps to fix  the codes resulted in incorrect answers. I will give you first the  problem description, then the code that resulted in incorrect answers.   You will fix the Python code, just return the Python code block   itself without explanations, try not to change the code structure. Remember, do not give any explanations. Here is the   problem description:  $<$description$>$   Here is the code that   resulted in wrong answers: $<$code$>$   \\ \hline 
 
  \end{tabular}
  \caption{  Prompts used to generate code with Gemini. We used the same prompts with CodeLlama and Codestral with a slight difference: The first parts that define the general task is given as system prompt while the rest is given as user prompt.} 
  \label{Tab_prompts}
\end{table*}

\section{Our Dataset}\label{sec_data}

In this section, we share the process of collecting codes written by humans, the stages of code generation by large language models, the operations applied to the obtained data, and the statistics regarding the dataset.

% idaliu = idialu2024whodunit
% oedingen2024chatgpt = 
%hung = pan2024assessing
% suh = suh2024empirical
%pham = pham2024magecode
%vision = xu2025codevision
%guri = gurioli2024you
%wang = wang2023evaluating
% xu = xu2024detecting

\subsection{Collecting Human Written Codes}

The first step in creating the dataset is to identify coding problems. In this study, we used the CodeNet dataset\footnote{https://developer.ibm.com/exchanges/data/all/project-codenet/} prepared by IBM and previously used in other studies\cite{xu2025codevision, xu2024detecting}. CodeNet is a dataset that includes 4,000 different coding problems and solutions written by humans in different programming languages such as Python, C++, Java, Ruby. Sometimes, human-written codes in CodeNet pass all tests successfully, while sometimes they result in compile-time errors, runtime errors, time limit exceeds, or incorrect output errors.

In this study, we used the problems in CodeNet's 800-problem subset called "Python Benchmark". This subset includes 800 problems that only contain human-written codes that have passed all tests successfully.
While examining the problem descriptions, we noticed that most of them were assigned a score (in description text). Due to time and budget constraints, we took a sample of these problems, taking into account the scores in the problem descriptions. Specifically, to ensure as even a distribution as possible between the scores, we first sampled 60 problems in each of the first five groups (score ranges 100, 200, 300, 400, 500), making a total of 300 problems (5x60). Since there were only 17 problems with higher scores (600 and 700), we included these problems directly in the dataset, resulting in a final number of 317 problems.

All human-written code in CodeNet has labels indicating its status:
i) Accepted, ii) Runtime Error, iii) Wrong Answer, iv) Compile-time Error, and v) Time-Limit Exceed error. To increase the diversity in our dataset, we collected human-written codes with different labels for each coding problem we pulled from the "Python Benchmark" section using CodeNet's "Full (Original) Dataset" and "Full Dataset, Metadata Only" dataset. Specifically, we collected five code samples from each of the codes with accepted labels, runtime errors, and wrong outputs, creating a dataset of 4,755 (317 x 5 x 3 = 4,755) human-written codes in total. Since Python is an interpreted language, there is no code with the compile error tag. Furthermore, since the code with timeout errors is rare for the selected problem set, we did not include the code with these errors in our dataset.

\subsection{Generating  Code Snippets}

We used three different large language models to generate the codes for the selected problem set by the AI: i) CodeLlama \cite{roziere2023code} with 34 billion parameters, ii) Codestral \cite{mistral} with 22 billion parameters, iii) Gemini 1.5 Flash \cite{team2024gemini}. For each problem, we asked these models to write code in Python programming language. As mentioned above, we obtained human-generated code in 3 different cases. For each case, we generated three different codes. Specifically, we asked the models to generate code from scratch by giving only the problem description as the correspond to the accepted codes. In response to the codes containing runtime errors, we randomly selected one of the human-written codes containing both the problem description and the runtime error and asked the models to fix that code. Finally, in response to codes ending with an incorrect answer, we gave the models the problem description and a random selection of human-generated code in that case to fix. With this method, we obtained a total of 2,853 (317 x 3 x 3) AI-generated code samples. The commands given to the AI models are presented in \textbf{Table \ref{Tab_prompts}}. The original problem definitions were preserved in the commands. Only for the purpose of improving readability, multiple consecutive '\textbackslash n' characters in the original problem definitions were changed to a single '\textbackslash n'.

\subsection{Filtering}
After generating the codes using the models, we performed a quality control check to ensure their validity. Several issues were identified in some of the generated code snippets, including: (i) failure to produce any output, (ii) inclusion of code written in C-family languages instead of Python, and (iii) presence of meaningless characters, sentences, numbers, or dots. These problematic snippets were excluded, reducing the total number of AI-generated code samples to 2,828.
 
Furthermore, although the prompts specifically requested only code as output, some responses included additional explanations, either as standalone text or embedded within the code as comments. To ensure the dataset accurately reflects the intended challenge of detecting AI-generated code, we manually removed any explanations provided above or below the code. However, comments embedded within the code blocks were retained, provided they were appropriately marked as comment lines.

\subsection{Features \& Statistics}

To offer a deeper understanding of our dataset, we provide sample code for the counting trailing zero problem. Specifically, the problem is as follows.

\textit{For an integer n not less than 0, let us define f(n) as follows: }\\
$f(n) = 1 (if n < 2) $ \\
$ f(n) = n f(n-2) (if n \geq 2)  $ \\
\textit{Given is an integer N. Find the number of trailing zeros in the decimal notation of f(N).}

\textbf{Table \ref{Tab_code}} provides code samples in our dataset for the problem for all LLMs we use for the three different usage scenarios. For each scenario, we also provide an example human written code with a the corresponding status.

\begin{table*}[!htb]
\small
\centering
\setlength\tabcolsep{1.5pt}
\renewcommand{\arraystretch}{0.7}
  \begin{tabular}{ | p{0.3cm} |  p{3.6cm} || p{3.6cm} | p{2.95cm} |  p{4.7cm} | }
    \hline
   & \textbf{Human-Written} & \textbf{Gemini} & \textbf{Codestral} & \textbf{LLama}  \\ \hline
\rotatebox[origin=r]{90}{Generated from Problem Definition} & \begin{lstlisting}
n = int(input())
if n % 2: print(0); exit()
ans = 0
x = 10
while x <= 10**18:
    ans += n // x
    x *= 5
print(ans)
 \end{lstlisting}& 
\begin{lstlisting}
def count_trailing_zeros(n):
    if n < 2:
        return 0
    count = 0
    while n >= 5:
        n //= 5
        count += n
    return count

n = int(input())
print(count_trailing_zeros(n))
 \end{lstlisting}
 & 
\begin{lstlisting}
N = int(input())
count = 0
while N > 0:
    N //= 5
    count += N
print(count)
 \end{lstlisting}
& \begin{lstlisting}
def count_trailing_zeros(n):
  count = 0
  while n % 10 == 0:
    count += 1
    n //= 10
  return count
def f(n):
  if n < 2:
    return 1
  else:
    return n * f(n-2)
def main():
  n = int(input())
  result = count_trailing_zeros(f(n))
  print(result)
if __name__ == "__main__":
  main()
\end{lstlisting}
 \\ \hline
\rotatebox[origin=r]{90}{Fixing Runtime Error}
&\begin{lstlisting}
N=int(input())
if N%2==1:
    print(0)
else:
    ans=0
    div=10
    while N//div>0:
        ans+=N//div
        div*=5
print(ans)
 \end{lstlisting}
& \begin{lstlisting}
def f(n):
    if n < 2:
        return 1
    else:
        return n*f(n-2)
n = int(input())
t = f(n)
print(t)
i = 0
while t % 10 == 0:
    t = t // 10
    i += 1
print(i)
 \end{lstlisting}
 & 
\begin{lstlisting}
def f(n):
  if n < 2:
    return 1
  else:
    return n * f(n - 2)
n = int(input())
t = f(n)
i = 0
while t % 10 == 0:
    t //= 10
    i += 1
print(i)
 \end{lstlisting}
& \begin{lstlisting}
N = int(input())
if N % 2 == 1:
    print(0)
else:
    ans = 0
    div = 10
    while N // div > 0:
        ans += N // div
        div *= 5
    print(ans)
\end{lstlisting}
 \\ \hline
\rotatebox[origin=r]{90}{Correcting Incorrect Output}
&\begin{lstlisting}
N = int(input())
if (N < 10) or (N % 2 != 0):
    print(0)
def calc(N):
    S = str(N)
    cnt = 0
    for x in S:
        if(x == "0"):
            cnt +=1
    return cnt
#print(N % 10)
number = 0
plus = 10
while N >= 10:
    if(N / 10 >= 10):
        number += plus
        plus *= 10
    N = N / 10
#print(N)
print(number+int(N))
 \end{lstlisting}
&  \begin{lstlisting}
N = int(input())
if (N < 10) or (N % 2 != 0):
    print(0)

def calc(N):
    S = str(N)
    cnt = 0
    for x in S:
        if(x == "0"):
            cnt +=1
    return cnt
#print(N % 10)
number = 0
plus = 10
while N >= 10:
    if(N / 10 >= 10):
        number += plus
        plus *= 10
    N = N // 10
#print(N)
print(number+int(N))
\end{lstlisting}
& \begin{lstlisting} n=int(input())
ans=0
n_5=0
n_2=0
if n%2==0:
  for i in range(1,19):
    ans+=n//(10**i)
x=n//10
for i in range(1,26):
  n_5+=x//(5**i)
for i in range(1,18):
  n_2+=x//(2**i)
ans+=min(n_5,n_2)
print(ans)
\end{lstlisting}
&
\begin{lstlisting}
N = int(input())
if (N < 10) or (N % 2 != 0):
    print(0)

def calc(N):
    S = str(N)
    cnt = 0
    for x in S:
        if(x == "0"):
            cnt +=1
    return cnt

number = 0
plus = 10
while N >= 10:
    if(N // 10 >= 10):
        number += plus
        plus *= 10
    N = N // 10
print(number+int(N))
\end{lstlisting} \\ \hline
  \end{tabular}
  \caption{ Human written and AI generated codes for the counting trailing zeros problem. Human written codes include with three status: Accepted, Runtime error, and Wrong answer} 
  \label{Tab_code}
\end{table*}

\begin{figure}[!htb]
\centering
\includegraphics[width=0.5\textwidth]{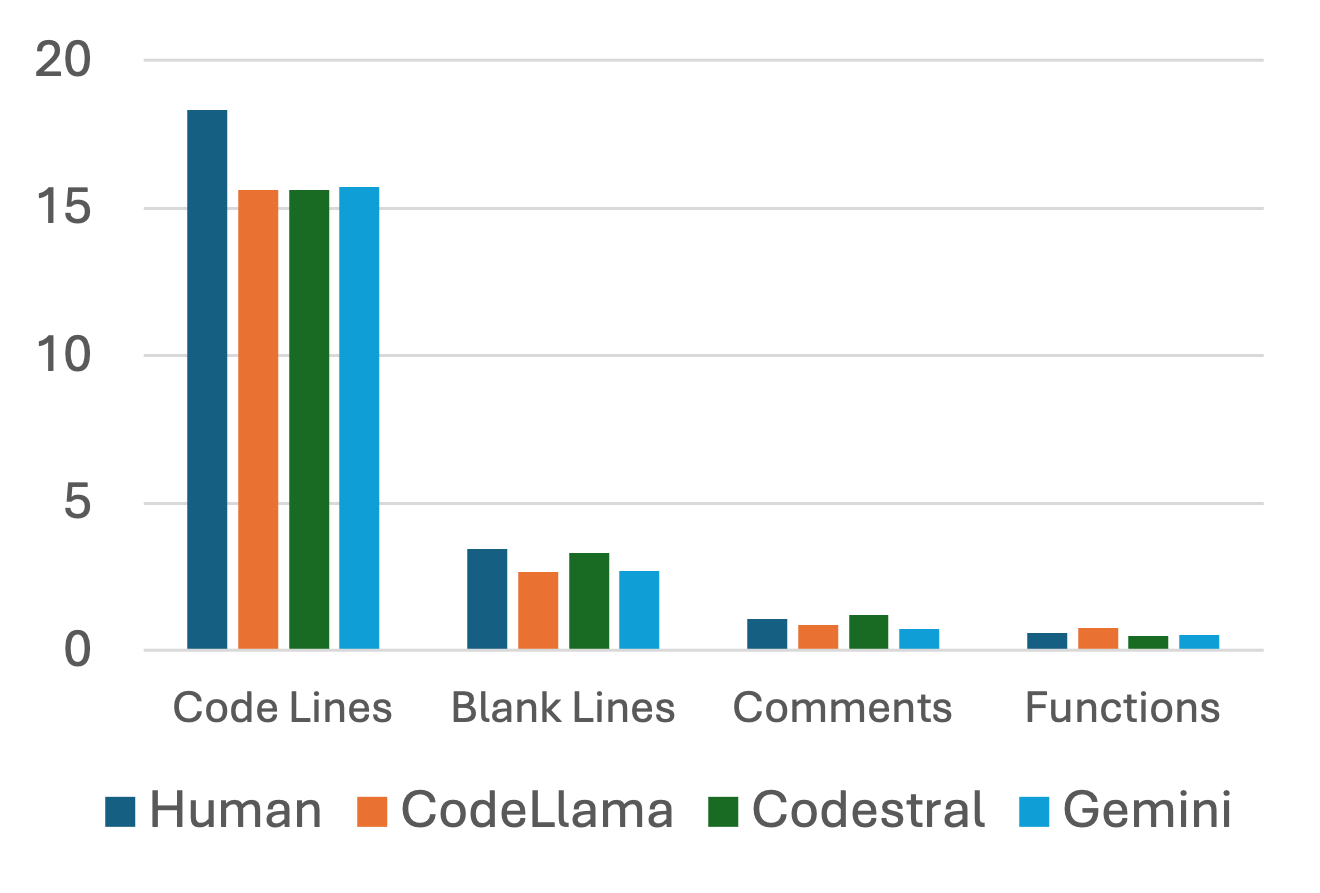}
\caption{Human written codes vs. Codes generated from scratch in AIGCodeSet.}
\label{fig:llm}
\end{figure}

To analyze the distinctions between AI-generated and human-written code, we calculated the average number of lines, blank lines, comments, and function definitions for human-written codes and for the outputs of the three LLMs separately. For AI-generated samples, we focused solely on codes generated from scratch to better capture distinctions in LLM-generated code. The results are presented in \textbf{ Figure \ref{fig:llm}}.

Human-written code is typically more extensive than code generated by LLMs. Among the LLMs, CodeLlama defines more functions than both humans and the other models. Codestral, on the other hand, uses blank lines and comments more frequently, aligning more closely with human-written code in this regard.

\section{Experiments}\label{sec_exp}
%\cite{idialu2024whodunit} Code stylometry, which examines unique coding styles that reveal patterns in how programmers write code

%\cite{suh2024empirical}:The study utilized SciTools Understand [13] to extract static source code metrics as features for training machine learning models, building on a foundation established in prior research [17, 28, 41, 44, 62]. Additionally, features identified by Aljehane et al. [18] were included, encompassing identifiers (method and variable names), Names and Operators in conditional statements (e.g., if, else, and while), as well as operators, keywords, arguments, and method signatures.

%Some studies, rely primarily on abstract syntax tree (AST) representations \cite{suh2024empirical}, in addition to AST \cite{10.1145/3626252.3630826} focusing on Java, also utilize Code2Vec, which is specifically trained on Java code.

In this section, we explain our experimental setup and present results for the baseline methods on our dataset. We also provide brief analysis on the impact of LLMs and how they are used in code generation on the performance of AI-generated code detectors.

\subsection{Experimental Setup}

We randomly sample 80\% of the dataset and use it for training while the rest is used for testing. We report $F_1$, precision, recall, and accuracy scores for models we implemented. %We also investigate the performance of models for various LLMs and code generation scenarios. 

%In our models, we represent codes using Ada embeddings and TF-IDF, similar to the approach in \cite{oedingen2024chatgpt}. 
We evaluate the performance of the following approaches.

\begin{itemize}
    \item \textbf{Ada Embeddings.} We represent the codes with Ada embeddings\footnote{\url{https://openai.com/index/new-and-improved-embedding-model/}} developed by OpenAI. We  train Random Forest (RF), XGBoost, and Support Vector Machine (SVM) models separately using our training set, yielding three different models.

    \item \textbf{TF/IDF Vector}. Following the approach of  \citet{oedingen2024chatgpt}, we represent the codes with TF/IDF vectors with a size of 5K, and train RF, XGBoost, and SVM models, separately.

    \item \textbf{Bayes Classifier}. \citet{oedingen2024chatgpt} report that Bayes classifier is highly effective in detecting AI generated code. We  use their implementation in our experiments.
\end{itemize}

\subsubsection{Experimental Results}
 
We assess the performance of the baseline models in our test set which includes all three LLMs and covers three different LLM usage scenarios. \textbf{Table \ref{Tab_res_all}} shows the results.

\begin{table}[!htb]
\centering
  \begin{tabular}{ | l |  c |c |c |c |}
    \hline
    Model & $F_1$ & Acc.   & P &   R \\ \hline
    RF w/ Ada  &  0.35	&   0.63	&   0.55	&   0.25 \\ \hline
    XGB w/ Ada  & 0.47	&   0.66	&   0.59	&   0.39  \\ \hline
    SVM w/ Ada  & 0.50	&   \textbf{0.71}	&   \textbf{0.79}	&   0.37 \\ \hline
    RF w/ TF-IDF &  0.32	&   0.57	&   0.43	&   0.25\\ \hline
    XGB w/ TF-IDF & 0.33	&   0.66	&   0.73	&   0.21 \\ \hline
    SVM w/ TF-IDF & 0.30	&   0.62	&   0.56	&   0.20 \\ \hline
    Bayes Classifier  &  \textbf{0.63}	&   0.63	&   0.52	&   \textbf{0.81} \\ \hline
  \end{tabular}
  \caption{The performance of baseline detection systems on the test set. The highest performing cases are written in \textbf{bold}. Acc: Accuracy, P: Precision, R: Recall} 
  \label{Tab_res_all}
\end{table}

Our observation are as follows. Firstly, Bayes classifier outperforms others in terms of $F_1$ and recall scores. However, its precision score is lower than others except RF model with TF-IDF vectors. Secondly, using Ada embeddings yield higher $F_1$ score compared to TF-IDF vectors. We achieve the highest accuracy and precision when we use SVM with Ada embeddings.

In order to explore the impact of LLM used for code generation, we split our test set based on LLMs and calculated the recall for each LLM. The results are shown in \textbf{Table \ref{Tab_res_llm}}

\begin{table}[!htb]
\centering
  \begin{tabular}{ | l |  c |c |c |}
    \hline
    Model & Gemini & Codestral & Llama \\ \hline
    RF w/ Ada  &  0.16 &  0.31 & 0.30\\ \hline
    XGB w/ Ada & 0.28 & 0.45 &  0.43\\ \hline
    SVM w/ Ada  & 0.21 &  0.47 & 0.43 \\ \hline
    RF w/ TF-IDF & 0.16  & 0.32 &  0.28 \\ \hline
    XGB w/ TF-IDF & 0.12 &  0.26  &  0.26 \\ \hline
    SVM w/ TF-IDF & 0.13 &  0.27 &  0.21\\ \hline
     Bayes Classifier  &  \textbf{0.74} & \textbf{0.88} & \textbf{0.81} \\ \hline
  \end{tabular}
  \caption{The percentage of correctly identified AI generated code samples for each LLM in the test set. The highest performing cases are written in \textbf{bold}. } 
  \label{Tab_res_llm}
\end{table}

We observe that the Bayes Classifier consistently achieves the highest recall across all cases. In addition, the Gemini model results in the lowest recall scores, indicating that it generates code that closely resembles human-written code. 

Lastly, we explore how the models' performance is affected by the code generation scenario. In particular, we divide the test based on the generation scenario and calculated the recall for each scenario. The results are shown is \textbf{Table \ref{Tab_res_scenario}}.

\begin{table}[!htb]
\centering
  \begin{tabular}{ | l |  p{1.2cm} |p{1.2cm} |p{1.2cm} |}
    \hline
    Model & Generate from Scratch & Fix Running Time Error & Correct Incorrect Output \\ \hline
    RF w/ Ada  & 0.552 & 0.129 & 0.087\\ \hline
    XGB w/ Ada  & 0.697  & 0.263 & 0.091\\ \hline
    SVM w/ Ada  & 0.701 & 0.220 & 0.048\\ \hline
    RF w/ TF-IDF & 0.532 & 0.134  & 0.197\\ \hline
    XGB w/ TF-IDF & 0.527 & 0.054 & 0.053\\ \hline
    SVM w/ TF-IDF & 0.493 & 0.059 & 0.183\\ \hline
    Bayes Classifier  & \textbf{0.985} & \textbf{0.677} & \textbf{0.769} \\ \hline
  \end{tabular}
  \caption{The percentage of correctly identified AI generated code samples for each LLM usage scenario in the test set. The highest performing cases are written in \textbf{bold}.} 
  \label{Tab_res_scenario}
\end{table}

We observe notable variation in recall scores across different generation scenarios. Specifically, all models achieve significantly higher recall scores when detecting code generated using problem definitions compared to other scenarios. This suggests that LLMs exhibit distinct coding styles. However, when prompted to fix a given code, the generated code closely resembles human-written code, as expected. We also observe this in the code samples provided in Table \ref{Tab_code}.

\section{Conclusion}\label{sec_conc}
In this study, we developed AIGCodeSet, a dataset for the AI-generated code detection problem, focusing on    Codestral, Codellama, and Gemini Flash 1.5 which are not well-studied in the literature. From the CodeNet dataset, we selected 317 problems and included five human submissions for each of the code status: Accepted, Runtime Error, and Wrong Answer, yielding 4,755 human-written code samples in total.  Next, we utilized the mentioned LLMs in three approaches: i) we used them to generate code based solely on the problem description filei ii) we employed them to fix human-written code with runtime errors using the problem description file, iii) we prompted them to correct human-written code that resulted in wrong answers, again with the problem description file. After filtering out improper outputs, we obtained a total of 2,828 AI generated code samples. %In addition, we shared the reasons for discarding certain outputs in our data link. 
Eventually,  AIGCodeSet consists of  7,583 (=4,755 + 2,828) data code samples. Additionally, we perform experiments on our dataset using baseline AI-generated code detection systems and present their results across various LLMs and usage scenarios.

\begin{comment}
    
Moving forward, we plan to enhance our dataset by incorporating a wider variety of programming languages, LLMs, and coding tasks to improve its utility for effective training and robust evaluation. Additionally, we aim to include more usage scenarios, such as hybrid code where LLMs generate only part of the code. Furthermore, we will conduct a user study involving students and software developers to understand how they utilize LLMs for code generation, helping us identify realistic LLM usage patterns.

\end{comment}

%the post-processing step, we were unable to verify whether the codes produced by the LLMs were complete, functional, or accurate. Specifically, we could not confirm if they generated the correct code or effectively fixed runtime errors or incorrect results in the sample codes, due to time constraints. We focused on whether the generated codes were written in Python, without considering their meaningfulness or accuracy. This can also be considered as potential future work on our dataset, focusing on evaluating the quality of LLMs or their ability to perform tasks and adapt based on prompts.

%\bibliographystyle{plainnat}
\bibliography{references}

\end{document}